\documentclass[10pt]{article}

\usepackage[margin=1.5in]{geometry} % full-width

\usepackage{amsmath}
\usepackage{amsthm}
\usepackage{amssymb}

\usepackage{wrapfig}

\usepackage{epsf}
\usepackage{epsfig}

\newcommand{\bfmX}{\boldsymbol{X}}
\newcommand{\bfmC}{\boldsymbol{C}}
\newcommand{\bfmZ}{\boldsymbol{Z}}

\newcommand{\bfX}{{\bf X}}

\newcommand{\bfzero}{{\bf 0}}

\usepackage[vlined,linesnumberedhidden,ruled,resetcount]{algorithm2e}

\newcommand{\bfmu}{\boldsymbol{\mu}}
\newcommand{\bfSigma}{\boldsymbol{\Sigma}}

\newcommand{\beginappendix}{
        \setcounter{table}{0}
        \renewcommand{\thetable}{A\arabic{table}}%
        \setcounter{figure}{0}
        \renewcommand{\thefigure}{A\arabic{figure}}%
        \setcounter{section}{0}
        \renewcommand{\thesection}{A\arabic{section}}%
        \setcounter{equation}{0}
     }

\title{Statistical disclosure control for numeric microdata via sequential joint probability preserving data shuffling}

\author{Elias Chaibub Neto}

\date{Sage Bionetworks, Seattle, WA 98121, USA \\ elias.chaibub.neto@sagebase.org}

\begin{document}

\maketitle

\begin{abstract}
Traditional perturbative statistical disclosure control (SDC) approaches such as microaggregation, noise addition, rank swapping, etc, perturb the data in an ``ad-hoc" way in the sense that while they manage to preserve some particular aspects of the data, they end up modifying others. Synthetic data approaches based on the fully conditional specification data synthesis paradigm, on the other hand, aim to generate new datasets that follow the same joint probability distribution as the original data. These synthetic data approaches, however, rely either on parametric statistical models, or non-parametric machine learning models, which need to fit well the original data in order to generate credible and useful synthetic data. Another important drawback is that they tend to perform better when the variables are synthesized in the correct causal order (i.e., in the same order as the true data generating process), which is often unknown in practice. To circumvent these issues, we propose a fully non-parametric and model free perturbative SDC approach that approximates the joint distribution of the original data via sequential applications of restricted permutations to the numerical microdata (where the restricted permutations are guided by the joint distribution of a discretized version of the data). Empirical comparisons against popular SDC approaches, using both real and simulated datasets, suggest that the proposed approach is competitive in terms of the trade-off between confidentiality and data utility. \\

\end{abstract}

\section{Introduction}

In this paper we address the problem of statistical disclosure control (SDC) for data publishing in the particular context of numerical microdata released in tabular format. (That is, where the data is organized into a table with the rows usually containing the data records and columns containing the attributes/variables measured for each data record.) Tabular datasets are widely used in most fields of science, and the current trend towards open and transparent science means that there is an ever increasing need for publishing tabular datasets supporting scientific research. Often times, the research involves privacy sensitive information, and these supporting datasets need to be masked (perturbed) in order to reduce disclosure risk before they can be shared publicly. Quite importantly, the amount of perturbation should ideally be enough to reduce the disclosure risk to an acceptable level, while retaining as much scientific utility as possible. (Business operations represent another case in point for the need of SDC for data publishing of tabular microdata.)

While there is a rich literature on SDC methodology for numeric microdata~\cite{Willenborg1996,hundepool2012}, most of the perturbation approaches are ``ad-hoc" in the sense that while they manage to preserve some particular aspects of the data, they end up modifying others. (E.g., while application of zero mean additive Gaussian noise to a variable approximately preserves its mean, it also increases its variance.)

Fully conditional specification (FCS) synthetic data approaches, on the other hand, aim to generate new datasets that follow the same joint probability distribution as the original data~\cite{Drechsler2011}. These approaches, however, rely on parametric statistical models, whose parameters need to be estimated from the data, or rely on non-parametric machine learning models, which need to fit well the data in order to generate credible and useful synthetic data. Another important drawback of these approaches is that they tend to perform better when the variables are synthesized in the correct causal order (i.e., in the same order as the true data generating process), which is often unknown in practice.

In this paper, we propose a fully non-parametric and model free data perturbation approach that circumvents important drawbacks of both perturbative and synthetic data based SDC methods. The approach, denoted sequential joint probability preserving data shuffling (SJPPDS), is based on sequential applications of restricted permutations to the numerical microdata, which are guided by the joint probability distribution of a discretized version of the data. As such, the approach is able to approximate the joint probability distribution of the original data without relying on parametric or non-parametric models, and without requiring any domain knowledge about the true data generating process. In practice, this means that the approach can be effective even when applied to small and noisy datasets, which tend to be challenging for synthetic data approaches that depend on models for fitting the data. Furthermore, in addition to approximately preserving the joint association structure of the data, the SJPPDS approach, by construction, produces a masked/perturbed dataset with the exact same marginal distributions as the original data. We implement two versions (full and simplified) of the SJPPDS method (described in Algorithms 1, 2, and 3 of Section 3).

Following~\cite{DomingoFerrerTorra2001}, we evaluate the tradeoff between data confidentiality and utility using a combination of multiple disclosure risk (DR) and information loss (IL) metrics. (Adoption of multiple metrics is generally advisable, since different metrics capture different aspects of the similarity between the original and masked datasets.) We compare the proposed approach against popular SDC approaches, using two real business microdata datasets, as well as, 60 simulated datasets. Our experimental results favor the SJPPDS approaches in terms of the trade-off of general data utility and disclosure risk.

\section{Related work}
%\vspace{-0.2cm}

Statistical disclosure control (SDC) is a mature field with many well established methods for the masking of numerical microdata~\cite{Willenborg1996,Drechsler2011,hundepool2012}. Traditional perturbative methods include: (i) microaggregation~\cite{DomingoFerrerMateoSanz2002,Templ2006}, where the original data records are first combined into small aggregate groups before their values are replaced by the mean of the aggregate groups; (ii) noise addition~\cite{Brand2004}, where masking is obtained by adding (independent or correlated) noise to the original data; and (iii) rank-swapping~\cite{Moore1996}, where each variable is first ranked in ascending order before each ranked value is swapped with another ranked value randomly chosen within a restricted range (such that the ranks of two swapped values cannot differ by more than a fixed percentage of the total number of records).

Synthetic data approaches have also been proposed for SDC, where instead of directly changing the original data values, completely new values are sampled from appropriate probability distributions that, hopefully, capture the essential statistical properties of the original data. For instance, under the assumption of multivariate normality, information preserving statistical obfuscation (IPSO) methods~\cite{Burridge2003,CanoTorra2009,Langsrud2019} aim to generate synthetic data that preserves the outputs of regression models (e.g., regression coefficients and/or covariance matrices).

Conceptually, the analytical properties of the original data are fully captured by the joint probability distribution (j.p.d.) of all variables in the dataset. The perturbative and synthetic data SDC approaches listed above, however, are not able to preserve the j.p.d. of the data. Fully conditional specification (FCS) methods~\cite{Drechsler2011}, on the other hand, aim exactly at simulating synthetic data with the same j.p.d. as the original data. The idea is to fully factorize the j.p.d. of the data, $P(X_1, X_2, X_3, \ldots, X_p)$, into a series of conditional distributions, $P(X_1 \mid X_2, \ldots, X_p) P(X_2 \mid X_3, \ldots, X_p) \ldots P(X_p)$, and then sequentially model and simulate one variable at a time, conditionally on the previous ones. The quality of the synthetic data generated using FCS methods depends, however, on the quality of the original data (as the conditional distribution models need to be estimated from the original data).  While parametric models can be used to estimate the conditional distributions, these models need to be chosen very carefully, and the FCS implementation based on the classification and regression tree (cart) model~\cite{Reiter2005} is often seen as the go to approach for FCS data synthesis, as it can flexibly model unusual data distributions and capture non-linear associations without requiring specification of the conditional distributions, and has been shown to provide the best empirical results in practice~\cite{DrechslerReiter2011,Nowok2016}. Despite these advantages, the quality of cart based data synthesis is still influenced by the order of the variables. While there is no standard way to select the variable order, the general recommendation is to try to emulate the causal ordering behind the data generating process giving rise to the original data. The problem, however, is that, in many applications this knowledge is not available.

\section{The proposed approach}

Before we describe the SJPPDS approach we first illustrate (using a toy example) how restricted permutations can be used as a data perturbation mechanism that (approximately) preserves the associations of numerical data. Consider the data from two highly correlated numerical variables, $X_1$ and $X_2$, in rows 1 and 2 of Table \ref{tab:1}, simulated from a bivariate normal distribution. Here we describe how to perform restricted permutations of the $X_1$ data relative to the $X_2$ variable.
%\vskip -0.05in
\begin{table}[!h]
\caption{Restricted permutation example}
  \label{tab:1}
  \centering
%\vskip -0.1in
\setlength\tabcolsep{4pt}
\begin{tabular}{l|rrrr|rrrr|rrrr}
\hline
$X_1$ & 8.87 &  9.57 &  9.61 &  9.36 &  9.75 & 10.51 & 10.01 &  9.67 & 10.29 & 11.42 & 12.11 & 11.64 \\
$X_2$ & 9.66 & 10.09 & 10.52 & 10.54 & 10.80 & 11.19 & 11.24 & 11.47 & 11.61 & 11.96 & 12.23 & 12.39 \\
$C_2$ & 1 & 1 & 1 & 1 & 2 & 2 & 2 & 2 & 3 & 3 & 3 & 3 \\
$X_1^\star$ & 9.61 &  9.36 &  8.87 &  9.57 &  9.75 & 10.51 &  9.67 & 10.01 & 11.64 & 12.11 & 10.29 & 11.42 \\
\hline
\end{tabular}
\end{table}

\begin{wrapfigure}{r}{0.55\textwidth}
\centerline{\includegraphics[width=\linewidth]{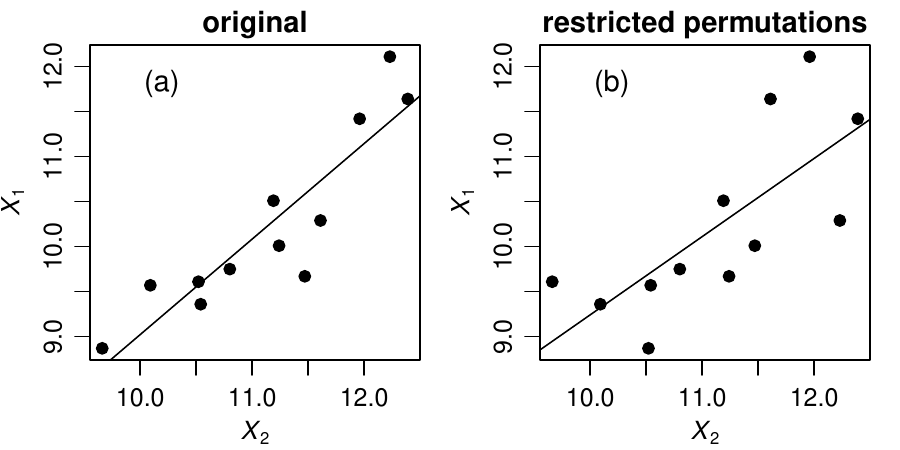}}
\vskip -0.1in
\caption{Restricted permutation example.}
\label{fig:restricted.perms.1}
%\vskip -0.1in
\end{wrapfigure}
The first step is to obtain a categorical version of the $X_2$ variable, denoted $C_2$, by discretizing the $X_2$ data into $n_c$ categories. In the example in Table \ref{tab:1} we discretize $X_2$ into $n_c = 3$ categories/levels, denoted as, ``1", ``2", and ``3" as shown in the $C_2$ row. (The discretization is performed by splitting the range of the $X_2$ data into 3 equally sized bins, and assigning the categorical labels ``1", ``2", and ``3" to the $X_2$ values that fall in each of these bins.) A restricted permutation of the $X_1$ data is then obtained by separately shuffling the values of the $X_1$ data within each level of the $C_2$ variable, as shown in the fourth row of Table \ref{tab:1}, where the shuffled $X_1$ values are denoted by $X_1^\star$. Figure \ref{fig:restricted.perms.1} plots the values $X_1$ against $X_2$ for the original data (panel a) and for the restricted permutation data (panel b). Clearly, the restricted permutation approach preserves well the association between the $X_1$ and $X_2$ variables. Observe, as well, that the marginal distribution of the shuffled data $X_1^\star$ is, by construction, identical to the marginal distribution of the original data $X_1$.

\begin{figure}[!h]
\centerline{\includegraphics[width=\linewidth]{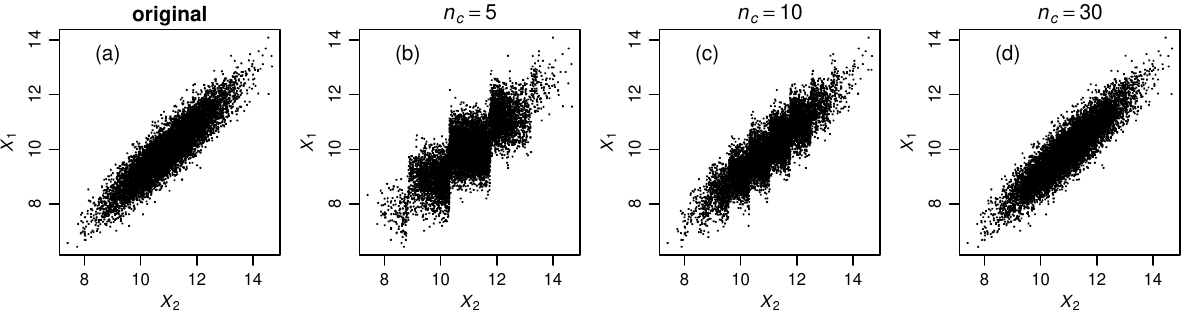}}
\vskip -0.1in
\caption{Effect of the number of categories/levels parameter, $n_c$, on the preservation of the association between $X_1$ and $X_2$ obtained by the restricted permutation approach. }
\label{fig:restricted.perms.2}
%\vskip -0.2in
\end{figure}

An important tuning parameter of the restricted permutation approach is the number of categories/levels, $n_c$, used in the discretization of $X_2$. The larger the $n_c$, the better is the association preservation, as illustrated in Figure \ref{fig:restricted.perms.2}, where we now simulate $n = 10,000$ records from  the same bivariate normal distribution as in Table \ref{tab:1} ($\bfmu = (10, 11)^T$ and $\sigma_{11} = 1$, $\sigma_{22} = 1$, and $\sigma_{12} = 0.9$).

On the other hand, the adoption of large values for the $n_c$ parameters will also lead to increased disclosure risk, as the larger the $n_c$, the smaller the overall amount of data shuffling (and in the extreme case when $n_c$ equals the number of records in the data, there is no shuffling of the data).

\subsection{Sequential joint probability preserving data shuffling}

We now describe how to perform (approximate) joint probability preserving data shuffling. The basic idea is to create a categorical version of the numeric microdata and then perform restricted permutations of the numeric data guided by the fully factorized joint probability distribution of the categorical data,
\begin{equation}
P(C_1, C_2, \ldots, C_p) = P(C_1 \mid C_2, \ldots, C_p) \, P(C_2 \mid C_3, \ldots, C_p) \, \ldots \, P(C_{p-1} \mid C_p) \, P(C_p)~.
\label{eq:categorical.factorization}
\end{equation}
Algorithm \ref{alg:1} describes the approach in detail. (Note that it can be seen as a fully non-parametric, model free, and perturbative analog of the fully conditional specification data synthesis paradigm.)

\begin{algorithm}[!h]
\caption{Joint probability preserving data shuffling - full version (JPPDS-f)}\label{alg:1}
\KwData{Microdata, $\bfmX$; categorical version of the microdata, $\bfmC$}
\ShowLn Find the number of attributes, $p$, of $\bfmX$, i.e., $p \leftarrow \mbox{NumberOfColumns}(\bfmX)$ \\
\ShowLn Find the number of records, $n$, of $\bfmX$, i.e., $n \leftarrow \mbox{NumberOfRows}(\bfmX)$ \\
%\ShowLn  $idx_{rows} \leftarrow (1, 2, \ldots, n)$ \\
\ShowLn Initialize the matrix with shuffled data with the original microdata, i.e., $\bfmX_s \leftarrow \bfmX$ \\
\ShowLn \For{$i$ in 1, \ldots, p-1} {
  %\ShowLn $idx_{columns} \leftarrow (i + 1, \ldots, p)$ \\
  \ShowLn Create a new matrix, $\bfmC^{\ast}$, by selecting columns of $\bfmC$ ranging from $i+1$ until $p$, i.e.,  $\bfmC^\ast \leftarrow \bfmC[ , (i+1):p]$ \\
  \ShowLn Create a vector of length $n$ containing the character strings obtained by pasting together the columns of $\bfmC^\ast$. This vector contains the combinations of the categorical variables in $\bfmC^\ast$, and is denoted \textit{lcombs}. Note that some of the combinations in this vector can be repeats \\
  %\ShowLn Let \textit{ucombs} represent the vector containing only the unique combinations of the \textit{lcombs} vector, i.e., $ucombs \leftarrow \mbox{Unique}(lcombs)$. \\
  \ShowLn Get the vector containing only the unique combinations of the \textit{lcombs} vector, i.e., $ucombs \leftarrow \mbox{Unique}(lcombs)$ \\
  \ShowLn Get the length of the $ucombs$ vector, i.e., $n_u \leftarrow \mbox{Length}(ucombs)$ \\
  \ShowLn \For{$j$ in 1, \ldots, $n_{u}$} {
    \ShowLn Find the indexes of the rows of $\bfmC^\ast$ that have the variable combination in $ucombs[j]$, i.e., $idx \leftarrow \mbox{Which}(lcombs == ucombs[j])$ \\
    \ShowLn Obtain a randomly shuffled version of the indexes in $idx$, i.e., $idx_{s} \leftarrow \mbox{Shuffle}(idx)$ \\
    \ShowLn For columns ranging from 1 to $i$, replace the data in the rows of $\bfmX_s$ indexed by $idx$ by the data in rows $idx_s$ of $\bfmX$, i.e., $\bfmX_s[idx, 1:i] \leftarrow \bfmX[idx_s, 1:i]$ \\
  }
}
\ShowLn Randomly shuffle the rows of $\bfmX_s$, i.e., $\bfmX_s \leftarrow \bfmX_s[\mbox{Shuffle}(1:n),]$. \\
\KwResult{Return the shuffled dataset $\bfmX_s$}
\end{algorithm}

To fix ideas, suppose that the numerical microdata, $\bfmX$, contains 4 attributes, $X_1$, $X_2$, $X_3$, and $X_4$, each of which is discretized into $n_c = 10$ classes (named as ``1", ``2", \ldots, ``10") in order to produce the associated categorical variables $C_1$, $C_2$, $C_3$, and $C_4$. Suppose, as well, that the microdata contains 10,000 records, so that both the numeric microdata and its categorical version, $\bfmC$, have dimension $n = 10,000$ by $p = 4$. Now, consider the full factorization of the joint probability distribution of the categorical data,
\begin{equation}
P(C_1, C_2, C_3, C_4) = P(C_1 \mid C_2, C_3, C_4) \, P(C_2 \mid C_3, C_4) \, P(C_3 \mid C_4) \, P(C_4)~.
\label{eq:categorical.factorization}
\end{equation}
Application of algorithm \ref{alg:1} to the numeric microdata goes as follows. The outer \textit{for-loop} (starting at line 4) captures the separate terms of equation (\ref{eq:categorical.factorization}). For $i = 1$, the algorithm deals with the $P(C_1 \mid C_2, C_3, C_4)$ term. Line 5 creates the $\bfmC^\ast$ matrix by selecting columns 2, 3, and 4 of $\bfmC$. Line 6 creates the vector of combinations of the categorical variables in $\bfmC^\ast$ ($lcombs$) by pasting together the levels of the $C_2$, $C_3$, and $C_4$ variables into a character string. (For example, if the values of $C_2$, $C_3$, and $C_4$ for the first record in the data are given, respectively, by ``9", ``2", and ``6", then the value at the first position of $lcombs$ is given by the string $``9.2.6"$. Note that for $n_c = 10$ there are at most $10 \times 10 \times 10 = 1,000$ possible distinct combinations of levels.) Line 7 obtains the unique combinations/strings observed in the data ($ucombs$), and line 8 counts the number of unique combination ($n_u$) in the data. The inner \textit{for-loop} of algorithm \ref{alg:1} (starting at line 9) effectively performs a restricted permutation of $X_1$ relative to the combination of the $C_2$, $C_3$, and $C_4$ variables and generates the shuffled numeric data as follows. For each one of the unique combinations in $ucombs$, line 10 finds the indexes of the records that share that combination, and lines 11 and 12 shuffle the numerical microdata values of variable $X_1$ within these indexes. After running through all unique combinations the result is a dataset where the values of $X_1$ have been shuffled in a way that preserves (approximately) their association with the $\{ X_2, X_3, X_4 \}$ variables.

For $i = 2$, the algorithm deals with the $P(C_2 \mid C_3, C_4)$ term. Now, line 5 creates the $\bfmC^\ast$ matrix by selecting columns 3 and 4 of $\bfmC$. Line 6 pastes together variables $C_3$ and $C_4$, and now there are at most $10 \times 10 = 100$ possible combinations in $ucombs$. Quite importantly, note that after the algorithm selects the indexes of the records that will be shuffled (lines 10 and 11), rather than shuffling the $X_2$ values alone, it shuffles both the $X_2$ and $X_1$ values together, as described in line 12. This is done to preserve the association between all four variables, rather than just the association between $X_2$ and $\{ X_3, X_4 \}$.

For $i = 3$, the algorithm deals with the $P(C_3 \mid C_4)$ term. Now, $\bfmC^\ast$ corresponds to column 4 alone, and there are only 10 possible combinations in $ucombs$. For each level of the categorical variable $C_4$, after the algorithm selects the indexes of the records that will be shuffled (lines 10 and 11), it shuffles both the $X_3$, $X_2$ and $X_1$ values together (line 12) to preserve the associations between all four variables.

Finally, note that line 13 of Algorithm \ref{alg:1} performs one last random shuffle of the rows of the matrix $\bfmX_s$ (i.e., randomly shuffle the records of $\bfmX_s$), because after the completion of steps 1 to 12, the last column of $\bfmX_s$ is still identical to the last column of $\bfmX$.

It is important to point out that the amount of shuffling performed by Algorithm \ref{alg:1} increases as $i$ increases from 1 to $p-1$ (line 4), simply because the number of possible combinations of the levels of the $C$ variables in the $ucombs$ vector, $n_u$, decreases as the number of number of columns of the $\bfmC^\ast$ matrix decreases. (In the above toy example, $n_u \leq 1000$ in the first iteration, $n_u \leq 100$ in the second, and $n_u \leq 10$ in the third.) Note that in situations where the number of records is smaller than the number of combinations in the $ucombs$ vector, we have that the number of records sharing the same combination of categorical variables in $ucombs$ will be 1 in most cases, so that Algorithm \ref{alg:1} barely performs any shuffling during its first iterations.

This observation suggests that we might be able to simplify Algorithm \ref{alg:1} to improve its computational efficiency without sacrificing too much its performance. Algorithm \ref{alg:2} provides a simplified version where we remove the outer \textit{for-loop} and only shuffle together the data of the first $p-1$ variables within each level of the last variable. In this case, we are essentially leveraging the following factorization of the joint probability distribution of the categorical variables,
\begin{equation}
P(C_1, C_2, \ldots, C_{p-1}, C_p) = P(C_1, C_2, \ldots, C_{p-1} \mid C_p) \, P(C_p)~,
\label{eq:categorical.factorization.simplified}
\end{equation}
to guide the restricted permutations of the data.

\begin{algorithm}[!h]
\caption{Joint probability preserving data shuffling - simplified version (JPPDS-s)}\label{alg:2}
\KwData{Microdata, $\bfmX$; categorical version of the microdata, $\bfmC$}
\ShowLn Find the number of attributes, $p$, of $\bfmX$, i.e., $p \leftarrow \mbox{NumberOfColumns}(\bfmX)$ \\
\ShowLn Find the number of records, $n$, of $\bfmX$, i.e., $n \leftarrow \mbox{NumberOfRows}(\bfmX)$ \\
\ShowLn Initialize the matrix with shuffled data with the original microdata, i.e., $\bfmX_s \leftarrow \bfmX$ \\
\ShowLn Get the last column of $\bfmC$, i.e., $C_p \leftarrow \bfmC[, p]$ \\
\ShowLn Get the unique values of $C_p$, i.e., $C_{pu} \leftarrow \mbox{Unique}(C_p)$ \\
\ShowLn Get the length of the $C_{pu}$ vector, i.e., $n_u \leftarrow \mbox{Length}(C_{pu})$ \\
\ShowLn \For{$j$ in 1, \ldots, $n_{u}$} {
  \ShowLn Find the indexes of $C_p$ that have the value in $C_{pu}[j]$, i.e., $idx \leftarrow \mbox{Which}(C_p == C_{pu}[j])$ \\
  \ShowLn Obtain a randomly shuffled version of the indexes in $idx$, i.e., $idx_{s} \leftarrow \mbox{Shuffle}(idx)$ \\
  \ShowLn For the first $p - 1$ columns, replace the data in the rows of $\bfmX_s$ indexed by $idx$ by the data in rows $idx_s$ of $\bfmX$, i.e., $\bfmX_s[idx, 1:(p-1)] \leftarrow \bfmX[idx_s, 1:(p-1)]$ \\
}
\ShowLn Randomly shuffle the rows of $\bfmX_s$, i.e., $\bfmX_s \leftarrow \bfmX_s[\mbox{Shuffle}(1:n),]$. \\
\KwResult{Return the shuffled dataset $\bfmX_s$}
\end{algorithm}

It is important to point out, however, that because the data in the first $p-1$ variables are shuffled together by Algorithm \ref{alg:2}, we have that essentially only the data in the last column is shuffled relative to the first $p-1$ columns. Hence, in order to make sure that the data in all columns are shuffled relative to each other we actually implement a sequential version of Algorithm \ref{alg:2} where it is sequentially applied to shifting orderings of the columns of the data, as described in Algorithm \ref{alg:3}. (Note that it is also usually beneficial to use the sequential approach described in Algorithm \ref{alg:3} in conjunction with Algorithm \ref{alg:1}, since the full JPPDS approach implemented in Algorithm \ref{alg:1} can still fail to adequately shuffle the data in the first columns of the dataset.)

\begin{algorithm}[!h]
\caption{Sequential joint probability preserving data shuffling (SJPPDS)}\label{alg:3}
\KwData{Microdata, $\bfmX$; number of classes/levels, $n_{c}$}
\ShowLn Find the number of attributes, $p$, of $\bfmX$, i.e., $p \leftarrow \mbox{NumberOfColumns}(\bfmX)$ \\
\ShowLn Create the categorical version of $\bfmX$, denoted $\bfmC$, by discretizing each column of $\bfmX$ into $n_{c}$ categories: $\bfmC \leftarrow \mbox{CategorizeData}(\bfmX, n_{c})$ \\
\ShowLn Generated the masked data, $\bfmX^\star$, by using the JPPDS algorithm: $\bfmX^\star \leftarrow \mbox{JointProbabilityPreservingDataShuffling}(\bfmX, \bfmC)$ \\
\ShowLn \For{$i$ in 1, \ldots, p - 1} {
  \ShowLn Change the order of the columns of $\bfmX^\star$, so that the first column is placed last, i.e., $\bfmX^\star \leftarrow \bfmX^\star[, c(2:p, 1)]$ \\
  \ShowLn Update the categorical version of the masked microdata, i.e., $\bfmC^\star \leftarrow \mbox{CategorizeData}(\bfmX^\star, n_{c})$ \\
  \ShowLn Update the masked microdata, i.e., $\bfmX^\star \leftarrow \mbox{JointProbabilityPreservingDataShuffling}(\bfmX^\star, \bfmC^\star)$ \\
}
\ShowLn Restore column order to the order in the original dataset, i.e., $\bfmX^\star \leftarrow \bfmX^\star[, c(2:p, 1)]$ \\
\KwResult{Return the masked dataset $\bfmX^\star$}
\end{algorithm}

Going back to the toy example with four variables, we have that line 3 of Algorithm \ref{alg:3} shuffles $X_4$ relative to $\{ X_1, X_2, X_3 \}$, while the \textit{for-loop} in lines 4 to 7  sequentially apply the JPPDS (full or simplified) to the following data orderings of the data, $\{X_2, X_3, X_4, X_1 \}$, $\{X_3, X_4, X_1, X_2 \}$, and $\{X_4, X_1, X_2, X_3 \}$, so that all variables are shuffled relative to each other. (That is, $X_4$ is shuffled relative to $\{ X_1, X_2, X_3 \}$ in line 3; $X_1$ is shuffled relative to $\{ X_2, X_3, X_4 \}$ when $i = 1$ in the \textit{for-loop} starting in line 4; $X_2$ is shuffled relative to $\{ X_3, X_4, X_1 \}$ for $i = 2$; and $X_3$ is shuffled relative to $\{ X_4, X_1, X_2 \}$ for $i=3$.)

\section{Performance evaluation}

We compared the performance of the proposed method, in both real and simulated data experiments, against the following widely used data perturbation approaches in the SDC field: (i) microaggregation, implemented using three distinct grouping methods (namely, the maximum distance to average vector (mdav) method~\cite{DomingoFerrerMateoSanz2002}, the principal component analysis (pca) method, and the projection pursuit principal components (pppca) method~\cite{Templ2006}); (ii) noise addition~\cite{Brand2004} based on independent additive noise and correlated noise; and (iii) rank-swapping~\cite{Moore1996} based on the percentage of ranks to be swapped. All perturbation methods were implemented using the sdcMicro R package~\cite{sdcMicro2015}. Additionally, we compare the proposed methods against the cart approach for synthetic data generation~\cite{Reiter2005}, implemented with the synthpop R package~\cite{Nowok2016}. We did not pursue comparisons against IPSO methods because the assumption of multivariate normality across all variables made by these approaches is rather strong, and almost never realized in real settings (it is certainly violated in the two real datasets we evaluated).

In all experiments, we evaluated the tradeoff between data confidentiality and data utility using a combination of 3  disclosure risk and 7 information loss metrics. Disclosure risk (DR) metrics evaluate either re-identification disclosure risk (i.e., the risk that an intruder might be able to determine the subject/entity to whom a given masked data record belongs to) or attribute disclosure risk (e.g., the risk that an intruder can learn about the value of confidential variables). We adopt distance-based record linkage (DBRL)~\cite{PagliucaSeri1999,DomingoFerrerTorra2001} to quantify re-identification disclosure risk, and rank-based interval disclosure (RID)~\cite{DomingoFerrerTorra2001} and standard deviation-based interval disclosure (SDID)~\cite{MateoSanzDomingoSebeFerrer2004} to quantify attribute disclosure risk. (See Appendix A1.1 for a brief description of these metrics.)

Information loss (IL) metrics can be classified as either general or analysis specific measures~\cite{Woo2009}. General measures attempt to either directly measure statistical distances between the masked and original datasets, or to measure the closeness of specific distribution parameters (e.g., moments, quantiles, and other summary statistics). Analysis specific measures, on the other hand, compare the results from specific analysis performed in the original and masked datasets (e.g., compare the point estimates and confidence intervals of a regression analysis). Since our goal is to publish masked datasets supporting scientific research or business operations, which might be used by others in unanticipated ways, we adopted general utility metrics in our evaluations. The selected IL measures included the propensity-score based information loss (PS)~\cite{Woo2009}, the probabilistic information loss (PIL)~\cite{PilPaper2005}, and 4 distance-based IL metrics (namely, the mean absolute and mean squared error distances (MAE and MSE)~\cite{DomingoFerrerMateoSanzTorra2001,DomingoFerrerTorra2001b} and the bounded rank-based mean absolute and mean squared error distances (brMAE and brMSE)~\cite{ChaibubNeto2023}). Appendix A1.2 provides further details about these selected metrics.

Quite importantly, note that all 3 DR metrics (namely, DBRL, RID and SDID) and the 4 distance-based IL metrics (namely, MAE, MSE, brMAE, and brMSE) selected for our comparisons can only, strictly speaking, be applied to masking methods for which there exists a mapping between the original, $x_{ij}$, and masked values, $x_{ij}^\star$, and are not sensible choices for the comparison of synthetic data approaches (or methods that completely shuffle the data such as the SJPPDS approaches) where no such connection exists. Hence, in order to be able to use these metrics with the SJPPDS and cart approaches we resort to the computation of averaged sorted versions of these metrics, as described in Algorithm \ref{alg:averaged.sorted.metrics}.
\begin{algorithm}[!h]
\caption{Averaged sorted version of metric $M$. ($M$ can represent any of the DBRL, RID, SDID, MAE, MSE, brMAE, and brMSE metrics.)}\label{alg:averaged.sorted.metrics}
\KwData{Original microdata, $\bfmX$; masked microdata, $\bfmX^\star$}
%\ShowLn Find the number of variables (attributes), $p$, of $\bfmX$, i.e., $p \leftarrow \mbox{NumberOfColumns}(\bfmX)$ \\
\ShowLn \For{$j$ in 1, \ldots, p} {
  \ShowLn Find the order of variable $j$ in the original dataset, i.e., $o_j \leftarrow \mbox{Order}(\bfmX[, j])$, and sort the records of the original data (across all variables) according to $o_j$, i.e., $\bfmX_s \leftarrow \bfmX[o_j,]$  \\
  \ShowLn Find the order of variable $j$ in the masked dataset, i.e., $o_j^\star \leftarrow \mbox{Order}(\bfmX^\star[, j])$, and sort the records of the masked dataset (across all variables) according to $o_j^\star$, i.e., $\bfmX_s^\star \leftarrow \bfmX^\star[o_j^\star,]$  \\
  \ShowLn Compute the metric using the sorted datasets $\bfmX_s$ and $\bfmX_s^\star$, i.e., $M_j^s \leftarrow \mbox{Metric}(\bfmX_s, \bfmX_s^\star)$
}
\ShowLn Compute the average of the $M_j^s$ metrics, i.e., $\bar{M}^s \leftarrow p^{-1} \sum_{j=1}^{p} M_j^s$ \\
\KwResult{Return the averaged sorted metric, $\bar{M}^s$}
\end{algorithm}

The basic idea is to: (i) separately sort the original and masked datasets according to the value of one of their variables; (ii) compute the selected metric using the sorted datasets; and (iii) repeat steps (i) and (ii) for all variables, and compute and report the average metric. Intuitively, this procedure represents a sensible way to generate an approximate mapping between the masked and original values for methods for which such a mapping does not truly exists. Observe, as well, that for the SJPPDS and cart methods this procedure generates more conservative DR metrics (i.e., larger disclosure risks) than the direct calculation of the metrics in the unsorted data. In all comparisons, we use the average sorted versions of the DBRL, RID, SDID, MAE, MSE, brMAE, and brMSE metrics for the SJPPDS and cart methods, but the original version of these metrics for the remaining SDC approaches.

Following reference~\cite{DomingoFerrerTorra2001}, we combine all these disclosure risk and information loss metrics into a single formula (score) meant to capture the overall confidentiality/utility tradeoff and provide an overall ranking of these SDC methods. Our overall score formula is given by,
\begin{equation}
\mbox{overall score} = \left[\frac{\mbox{DBRL}}{6} + \frac{\mbox{RID}}{6} + \frac{\mbox{SDID}}{6} \right] + \left[\frac{\mbox{PS}}{6} + \frac{\mbox{PIL}}{6} + \frac{\mbox{brMAE}}{12} + \frac{\mbox{brMSE}}{12} \right]~,
\label{eq:overall.score}
\end{equation}
which gives equal weights for the combination of DR metrics (1/6 + 1/6 + 1/6 = 0.5) and combination of IL metrics (1/6 + 1/6 + 1/12 + 1/12 = 0.5). Because the brMAE and brMSE are similar metrics we give then a 1/12 weight, so that together they contribute the same weight as the other IL metrics. We do not include the MAE and MSE metrics in the computation of the score because they are unbounded. Because the PS metric range is between [0, 0.25] we multiply it by 4, so that it is also bounded between 0 and 1, as all the other metrics used for the computation of the overall score in (\ref{eq:overall.score}). Note that lower overall scores are preferred since they indicate lower disclosure risks and lower information losses.

For each of the masking methods, we used the above score formula to:
\begin{enumerate}
\item First select the best parameter across a grid of 30 tuning/perturbation parameter values (under the constraint that the DBRL metric is below a given fixed threshold).

\item Then to compare the overall performance of the SDC methods based on the selected best parameter score.
\end{enumerate}
Note that we only select tuning parameters that produce DBRL values below a given fixed threshold, because, in practice, we are not interested in masking methods that preserve well data utility at the expense of high disclosure risks. Figure \ref{fig:dbrl.thresholds} in the Appendix provides an illustrative example of the selection of best tuning/perturbation parameter values conditional on different DBRL thresholds. In our comparisons we adopt a DBRL threshold of 0.2. (We choose this relatively high threshold because the noise-addition methods generated high values of DBRL in the simulated datasets and choosing a lower threshold would lead to discarding these methods from the comparisons. In practice, however, one might be interested in lower threshold values.)

For the real data experiments, we use 2 business microdata datasets, Census and Tarragona, which have been traditionally used to evaluate SDC approaches for numerical microdata. Both datasets are available with the sdcMicro R package~\cite{sdcMicro2015}. (Note that the Census dataset is named as CASCrefmicrodata in the sdcMicro package.) The Census data is composed of 1080 records on 13 variables (but we only use 12 of the Census dataset variables for our illustrations because one of the variables, PEARNVAL, is a linear combination of the other 12 variables, what causes problems for the calculation of the propensity-score based information loss metric. In practice, we can always apply the masking methods to the 12 variables and simply compute the masked version of the PEARNVAL variable from the other variables.) The Tarrogana data is composed of 834 records on 13 variables. In addition to being relatively small, both datasets contain variables with very different scales. Furthermore, the variables in the Tarragona dataset show fairly skewed data distributions and contain outliers. For the simulated data illustrations, we performed 2 experiments simulating data from multivariate normal and correlated exponential variables. In each experiment, we simulate 30 distinct $\bfmX$ matrices of dimension $n = 500$ by $p = 10$ with different location, scale, and correlation strengths. (See Appendix A1.3 for details.)

For both the real and simulated data experiments, we adopted the following tuning (perturbation) parameter grids:
\begin{itemize}
\item Aggregation parameters in the range $\{2, 3, \ldots, 31 \}$ for the microaggregation methods.
\item Noise percentage parameter in the range $\{ 1, 5, 9, 13, \ldots, 117 \}$\% for the noise addition methods.
\item Swap percent parameter in the range  $\{ 2, 4, 6, \ldots, 60 \}$\% for rank-swapping.
\item Number of classes parameter ($n_c$) in the range $\{10, 20, \ldots, 300 \}$ for both the SJPPDS methods.
\item 30 distinct randomly drawn variable orderings for the cart approach.
\end{itemize}

\begin{figure}[!h]
\centerline{\includegraphics[width=\linewidth]{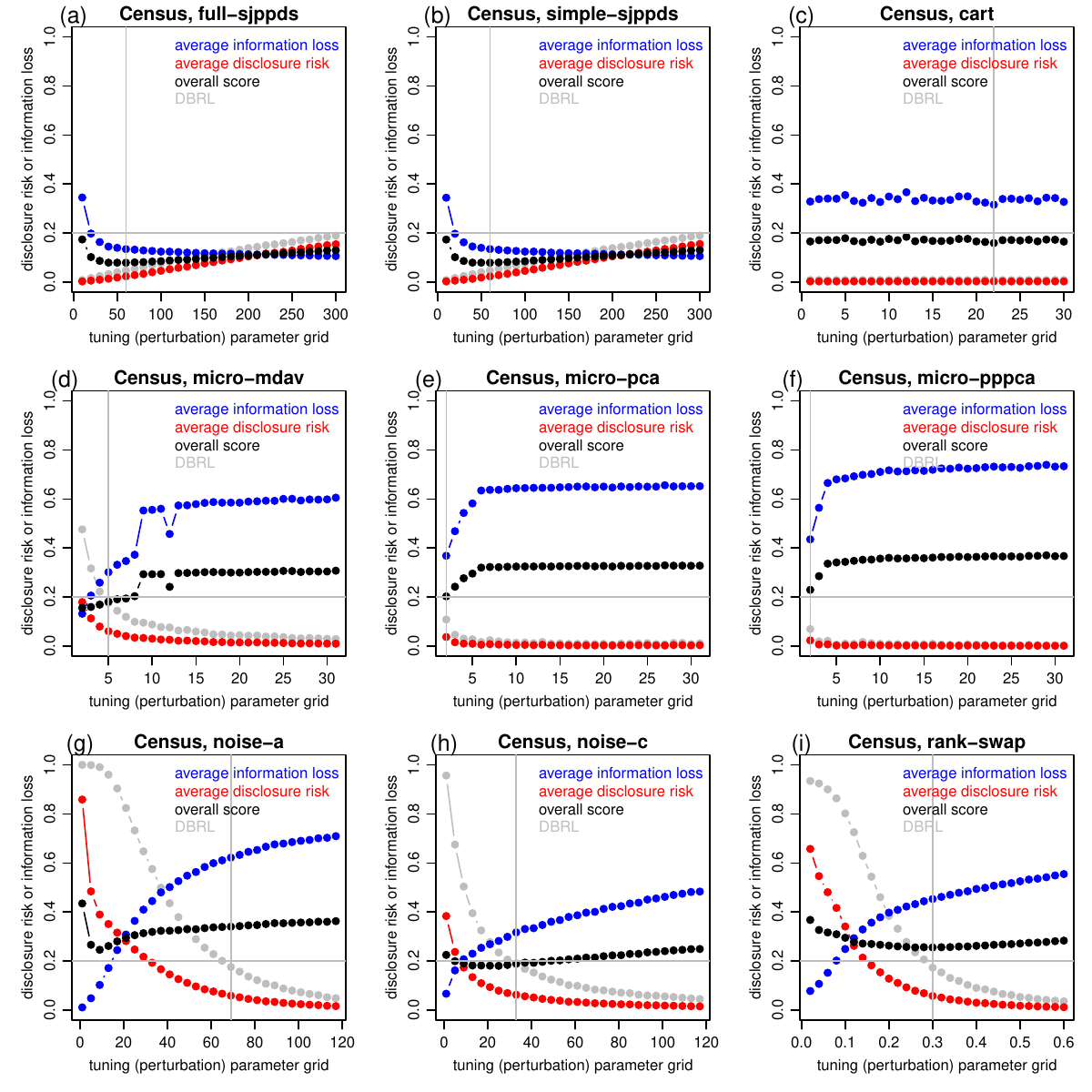}}
%\vskip -0.15in
\caption{Tradeoff between information loss and disclosure risk in the Census dataset.}
\label{fig:tradeoff.census}
%\vskip -0.1in
\end{figure}

Figure \ref{fig:tradeoff.census} reports the tradeoffs between information loss and disclosure risk across the grid of tuning/perturbation parameter values for each one of the SDC methods in the Census dataset. In all panels, the blue curve represents the average information loss score,
\begin{equation}
\mbox{average information loss} = \frac{\mbox{PS}}{3} + \frac{\mbox{PIL}}{3} + \frac{\mbox{brMAE}}{6} + \frac{\mbox{brMSE}}{6}~,
\label{eq:average.info.loss}
\end{equation}
 the red curve represents the average disclosure risk score,
\begin{equation}
\mbox{average disclosure risk} = \frac{\mbox{DBRL}}{3} + \frac{\mbox{RID}}{3} + \frac{\mbox{SDID}}{3}~,
\label{eq:average.disclosure.risk}
\end{equation}
the black curve represents the overall score in equation \ref{eq:overall.score} (i.e., the average of the average information loss and average disclosure risk scores), and the grey curve represents the DBRL metric. (Note that each point on each of these curves actually corresponds to the median value from the 30 experiment replications.) The grey horizontal line shows the selected DBRL threshold (i.e., 0.2), while the grey vertical line shows the selected best parameter value (under the constraint that DBRL is less than 0.2). Figure \ref{fig:tradeoff.tarragona} in the Appendix shows the analogous plot for the Tarragona data experiments. (Due to space limitations, we do present the condidentialy/utility tradeoff plots for the simulated datasets, as those would require a separate plot for each one of the 60 simulated datasets.)

\begin{figure}[!h]
\centerline{\includegraphics[width=\linewidth]{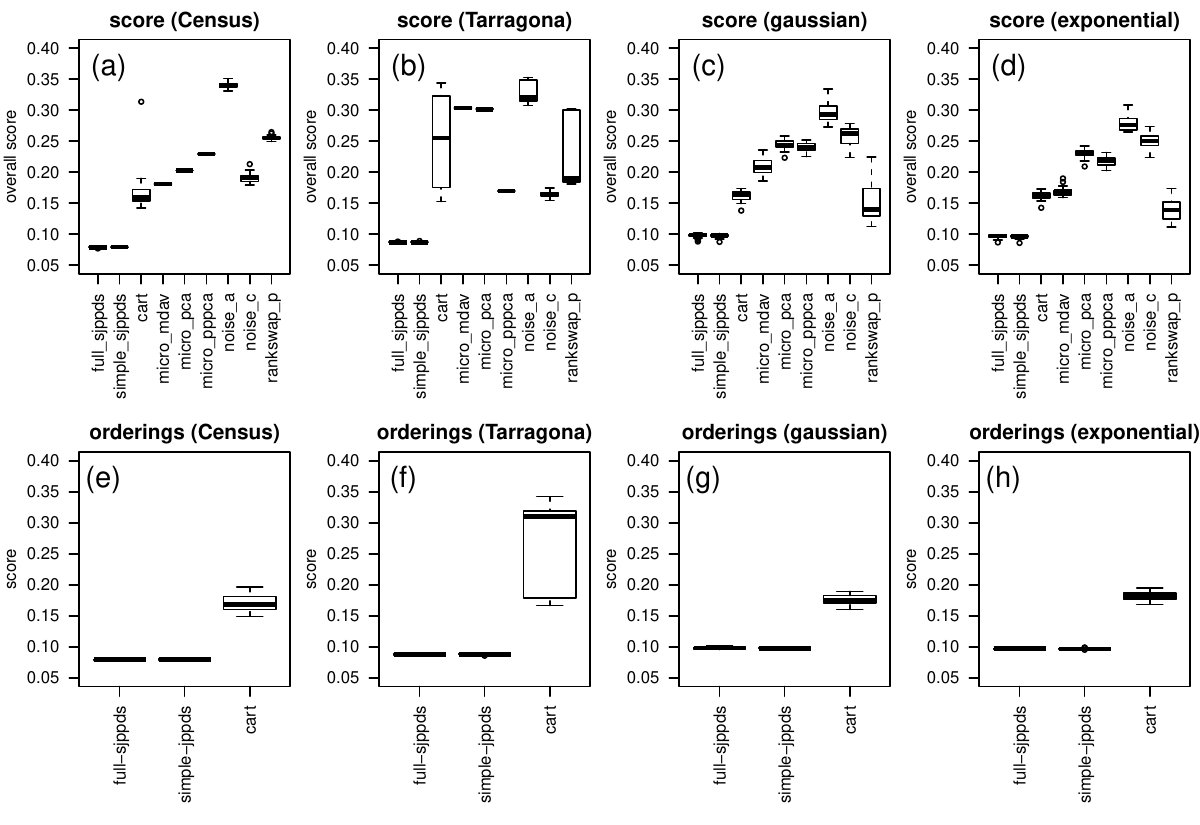}}
%\vskip -0.15in
\caption{Results from the real and simulated data experiments.}
\label{fig:experiments}
%\vskip -0.1in
\end{figure}

Figure \ref{fig:experiments} reports the results from all experiments. The top panels show boxplots comparing the overall scores of each of the SDC methods. For the real data experiments (panels a and b), the boxplots report the results from 30 replications of the experiments based the best selected tuning parameter (with the exception of the microaggregation methods, which are deterministic, and therefore based on a single experiment run). For the simulated data experiments (panels c and d) the boxplots report the scores from the 30 distinct simulated datasets. (Each method is first evaluated on a grid of 30 distinct tuning parameter values, but the boxplots only report the results based on the best selected tuning/perturbation parameter.) In terms of the data confidentiality/utility tradeoff measured by the overall score, the SJPPDS approaches outperformed all other methods in all experiments.

\begin{figure}[!h]
\centerline{\includegraphics[width=4.5in]{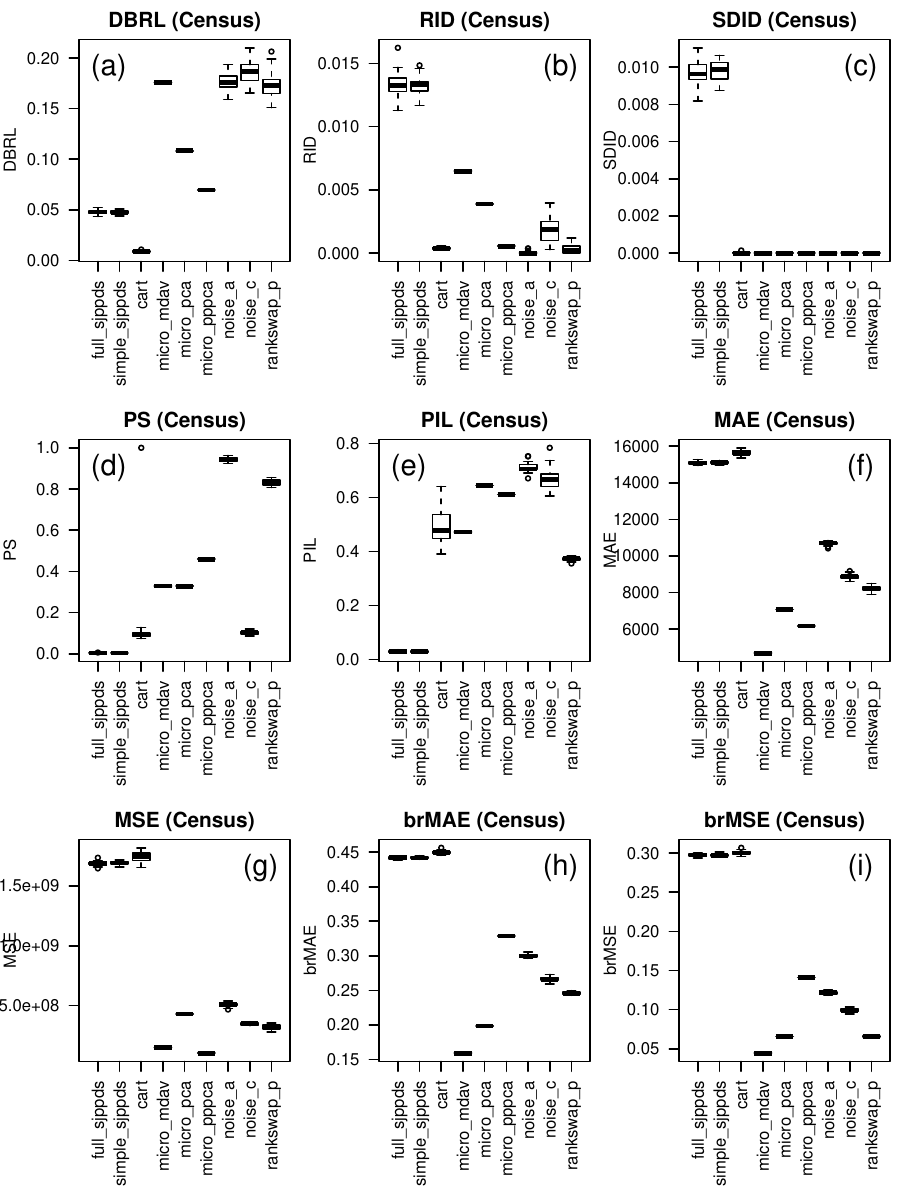}}
%\vskip -0.15in
\caption{Metrics comparison in the Census data experiment.}
\label{fig:metrics.real.data.census}
%\vskip -0.1in
\end{figure}

\begin{figure}[!h]
\centerline{\includegraphics[width=4.5in]{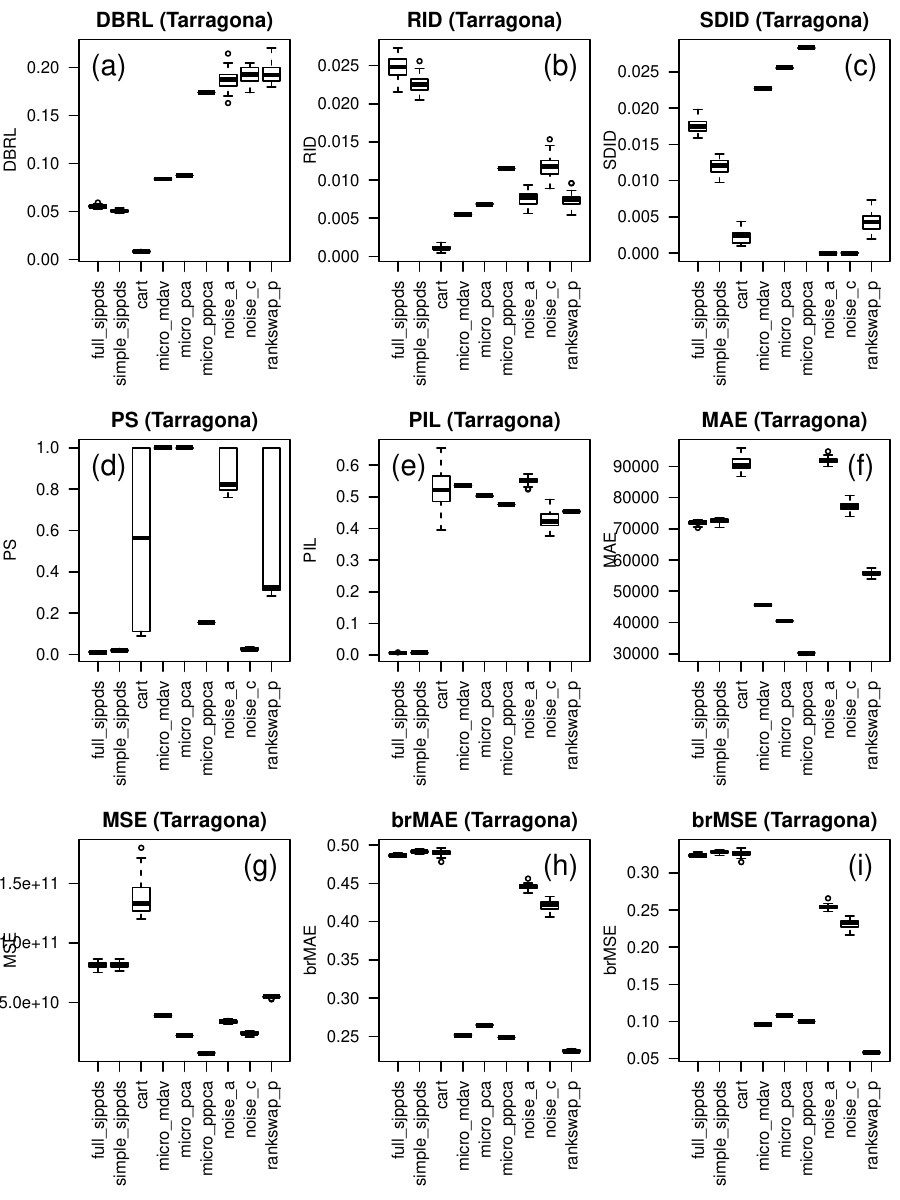}}
%\vskip -0.15in
\caption{Metrics comparison in the Tarragona data experiment.}
\label{fig:metrics.real.data.tarragona}
%\vskip -0.1in
\end{figure}

\begin{figure}[!h]
\centerline{\includegraphics[width=4.5in]{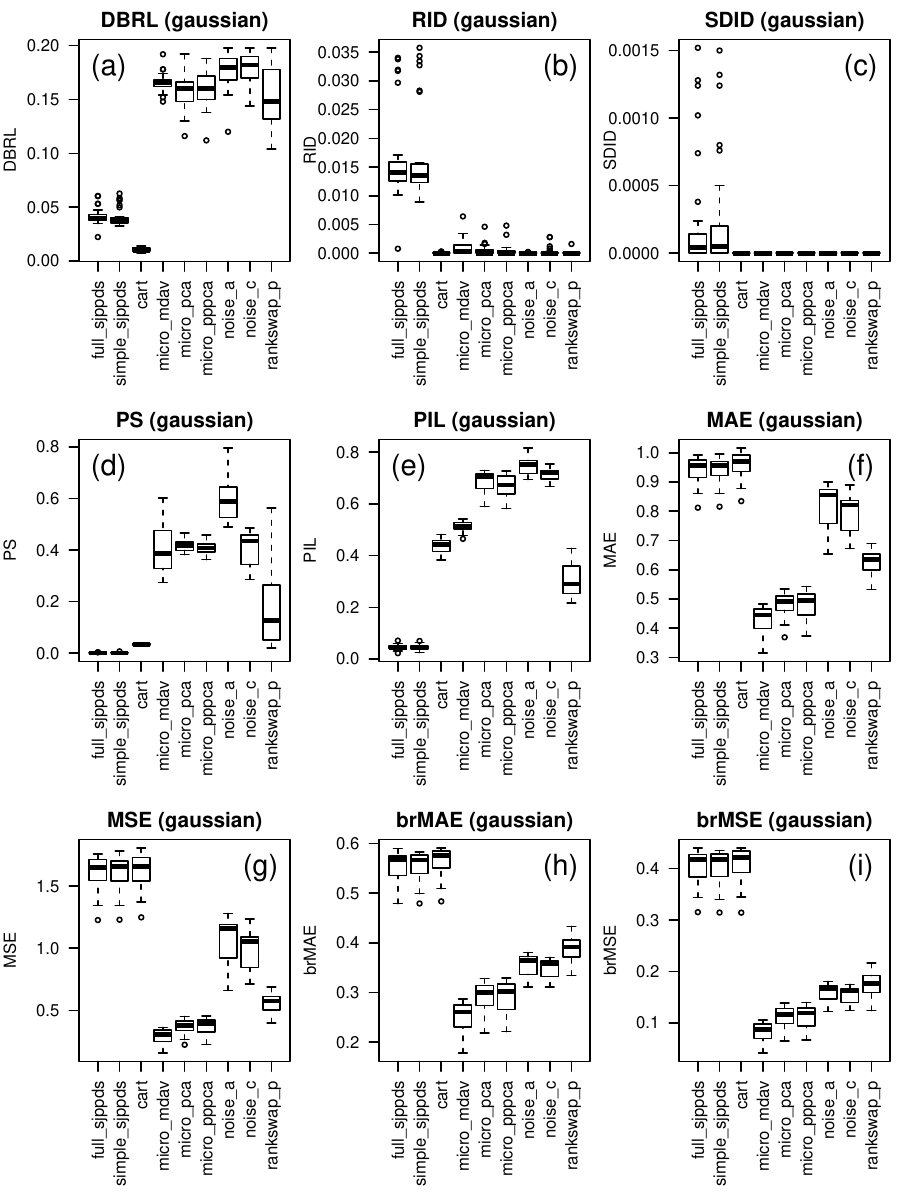}}
%\vskip -0.15in
\caption{Metrics comparison in the simulated gaussian data experiment.}
\label{fig:metrics.simulated.data.gaussian}
%\vskip -0.1in
\end{figure}

\begin{figure}[!h]
\centerline{\includegraphics[width=4.5in]{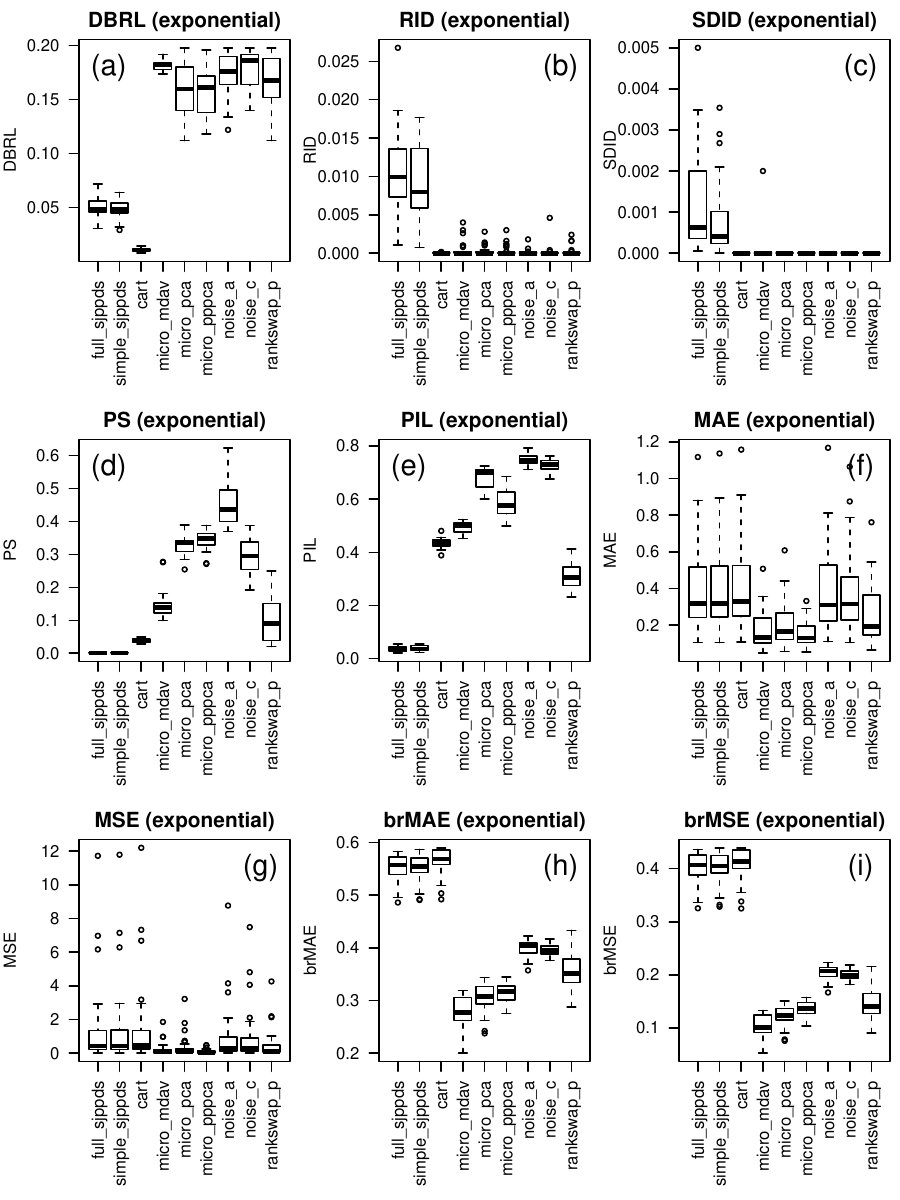}}
%\vskip -0.15in
\caption{Metrics comparison in the simulated exponential data experiment.}
\label{fig:metrics.simulated.data.exponential}
%\vskip -0.1in
\end{figure}

It is also interesting to notice that while the cart method performed relatively well in the Census and simulated datasets (it was outperformed only by the SJPPDS methods in panel a and by the SJPPDS and rank-swapping approaches in panels c and d), it's performance was considerably worse in the Tarragona data (panel b). A possible explanation is that the variables in the Tarragona dataset show fairly skewed distributions, which present a greater challenge for a model based approach as the cart method. Additionally, as pointed out before, an important drawback of data synthesis methods based on fully conditional specification (such as cart) is that the quality of the synthetic data depends on the order of the variables used for the data synthesis. The SJPPDS approaches, on the other hand, do not suffer from this caveat. To illustrate this point, the bottom panels of Figure \ref{fig:experiments} compare the overall scores of the cart model computed over 30 distinct orderings of the variables, against the SJPPDS approach computed using the same distinct orderings as the cart model (and using the best selected number of classes/labels tuning parameter). The experiments clearly show that the results from the SJPPDS approaches are much less variable (note the much narrower boxplots) than the results from the cart model.

Figures \ref{fig:metrics.real.data.census}, \ref{fig:metrics.real.data.tarragona}, \ref{fig:metrics.simulated.data.gaussian}, and \ref{fig:metrics.simulated.data.exponential} present more detailed descriptions of the experimental results looking at each of the DR and IL metrics separately for each one of experiments. The boxplots again report the metric values based on the best selected tuning/perturbation parameter across the 30 replications of the experiments. These figures show that no single SDC method outperformed all others across all DR and IL metrics.

In terms of DBRL, Figures \ref{fig:metrics.real.data.census}a, \ref{fig:metrics.real.data.tarragona}a, \ref{fig:metrics.simulated.data.gaussian}a, and \ref{fig:metrics.simulated.data.exponential}a show that the cart and SJPPDS methods outperformed all other methods. (As one would expect, the cart method, being a synthetic data approach, showed the lowest rates of record linkage.) In terms of RID, Figures \ref{fig:metrics.real.data.census}b, \ref{fig:metrics.real.data.tarragona}b, \ref{fig:metrics.simulated.data.gaussian}b, and \ref{fig:metrics.simulated.data.exponential}b show that the SJPPDS approaches produced higher disclosure risks than the other methods, although the disclosure risks were still fairly low (e.g., below 0.04 in all experiments). Similarly, for the SDID metric, Figures \ref{fig:metrics.real.data.census}c, \ref{fig:metrics.real.data.tarragona}c, \ref{fig:metrics.simulated.data.gaussian}c, and \ref{fig:metrics.simulated.data.exponential}c show that the SJPPDS approaches tended to produce higher disclosure risks than the other methods, although they outperformed the microaggregation methods on the Tarragona dataset (Figure \ref{fig:metrics.real.data.tarragona}c) and, again, tended to be fairly low (e.g., below 0.02 in all experiments).

In terms of IL metrics, panels d and e of Figures \ref{fig:metrics.real.data.census}, \ref{fig:metrics.real.data.tarragona}, \ref{fig:metrics.simulated.data.gaussian}, and \ref{fig:metrics.simulated.data.exponential} show that the SJPPDS methods outperformed all other methods in terms of the PS and PIL metrics, being considerably better w.r.t. the PIL metric. On the other hand, panels f, g, h and i of Figures \ref{fig:metrics.real.data.census}, \ref{fig:metrics.real.data.tarragona}, \ref{fig:metrics.simulated.data.gaussian}, and \ref{fig:metrics.simulated.data.exponential} show that the cart and SJPPDS methods tended to be outperformed by all other methods in terms of distance-based IL metrics.

Hence, all in all, the better overall performance of the SJPPDS approaches (Figure \ref{fig:experiments}) appears to be explained by their ability to trade a small increase in attribute disclosure risk (measured by RID and SDID) by considerable lower levels of information loss w.r.t. distribution parameters (measured by the PS and PIL), while still maintaining a relatively low re-identification disclosure risk (measured by DBRL), despite their increased information loss w.r.t. the distance-based IL metrics.

\section{Computation time benchmarking experiments}

We evaluated the computation time of the proposed SJPPDS approaches against all other SDC methods evaluated in this paper. Each experiment was replicated 30 times, and the user times computed by the \texttt{system.time} function of R base were recorded. All experiments were performed on a Windows machine with processor Intel(R) Core(TM) i7-7820HQ CPU \@ 2.90GHz 2.90 GHz and 64 GB of RAM. See Appendix A1.4 for further details about these experiments.

\begin{figure}[!h]
\centerline{\includegraphics[width=\linewidth]{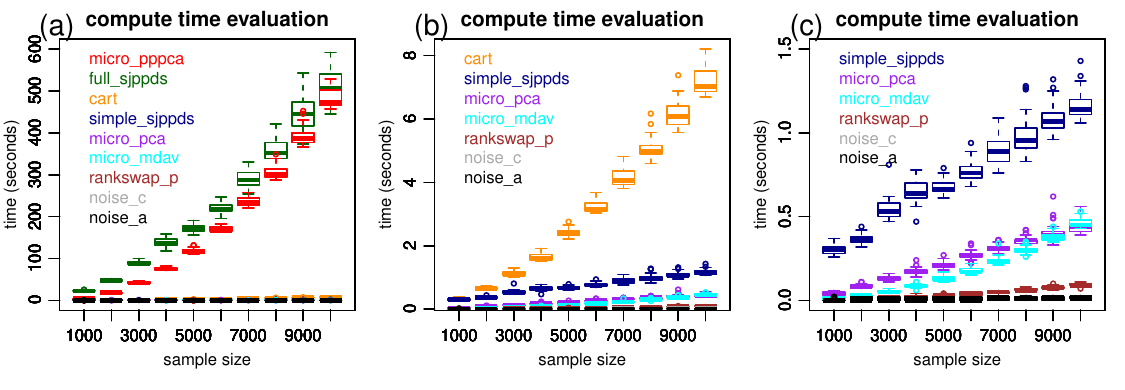}}
\vskip -0.15in
\caption{Computation time benchmarking of all SDC methods.}
\label{fig:computation.1}
%\vskip -0.2in
\end{figure}

In the first experiment (Figure \ref{fig:computation.1}) we evaluated user time for datasets with number of records increasing from 1,000 to 10,000 and with $p = 12$ variables. Panel a shows that the full SJPPDS and microaggregation using projection pursuit principal components methods were by far the most time expensive approaches. Panel b reports the results after removing these two methods, and shows that cart is the third most expensive approach. Finally, panel c reports the results after removing the 3 most expensive approaches, and shows that the simplified SJPPDS method was the forth most expensive one. Observe, however, that the approach is still able to mask a dataset containing 10,000 records and 12 variables in under 1.5 seconds. As one would expect, the noise addition methods are by far the fastest ones.

Even though we made no efforts to optimize our code for speed (it is implemented in R), overall, the simplified SJPPDS method showed competitive speed, being considerably faster than cart, but still slower than the methods implemented in the sdcMicro R package (which uses internal C++ implementations for computational efficiency). In fact, the most time consuming step in our experiments was the computation of the disclosure/information loss metrics, rather than the application of the masking methods.

These experiments also clearly illustrate that the simplified SJPPDS method should be the default choice in practice, since its computation can be orders of magnitude faster than the full version, and both approaches deliver comparable results in terms of masking performance (as illustrated by the experiment's results in the previous section).

To better understand the computational costs of the simplified SJPPDS approach we show (see Appendix A2) that the time complexity of Algorithm 3 (SJPPDS) when the joint probability preserving data shuffling is performed using the simplified version described in Algorithm 2 (JPPDS-s) is $O(n (p^2 + n_c p))$, where, as described before, $n$ represents the sample size (number of rows), $p$ represents the number of attributes (number of columns), and $n_c$ represents the number of classes/labels (number of bins) parameter used in the discretization of the numeric microdata. This means that the computation time: (i) scales quadratically with the number of columns of the dataset when the number of rows and number of bins is fixed; (ii) scales linearly with the number of rows of the dataset when the number of columns and bins are fixed; and (iii) scales linearly with increases in the number of bins when the number of rows and columns are fixed.

\begin{figure}[!h]
\centerline{\includegraphics[width=\linewidth]{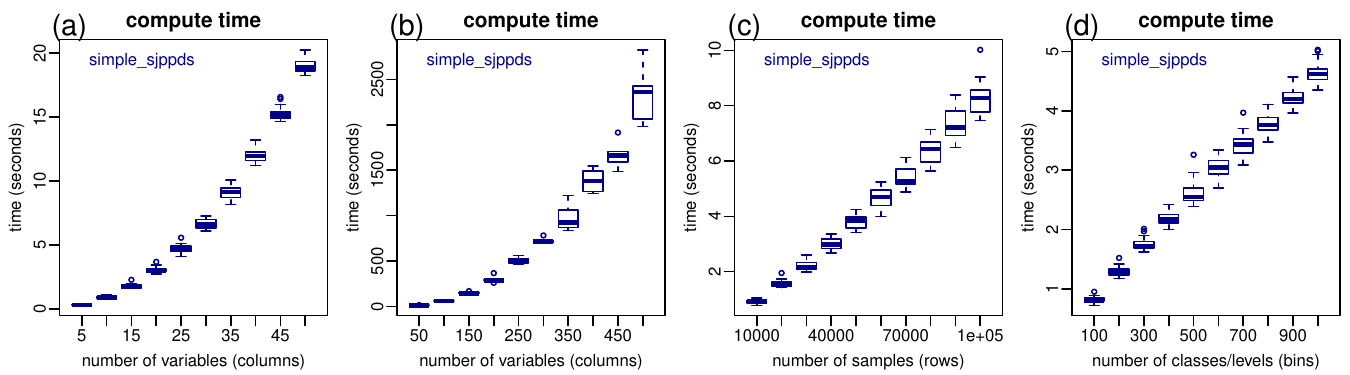}}
\vskip -0.15in
\caption{Compute time for the simplified JPPDS method.}
\label{fig:computation.2}
%\vskip -0.2in
\end{figure}

To illustrate each of these points we performed 3 additional experiments. In the first, we set $n_c$ to 100 and simulate datasets with $n = 10,000$ records and number of variables ($p$) increasing at first from 5 to 50 (Figure \ref{fig:computation.2}a) and then increasing from 50 to 500 (Figure \ref{fig:computation.2}b). These two panels clearly show a quadratic increase in computation time as $p$ increases. In the second experiment, we set $n_c$ to 100 and simulate datasets with $p = 10$ variables and sample sizes ($n$) increasing from 10,000 to 100,000 (Figure \ref{fig:computation.2}c), clearly illustrating the linear increase in computation time. Finally, in the third experiment we simulate datasets with $p = 10$ variables and $n = 10,000$ records and report the computation time for number of bins ($n_c$) increasing from 100 to 1,000, clearly illustrating that the computation time scales linearly with $n_c$.

\section{Final remarks}

We proposed a fully non-parametric and model free perturbative approach for SDC, that does not require any model specification nor is influenced by variable order. The method preserves the association structure of the data while generating marginal distributions that are identical to the original data. It is straightforward to implement and computationally light. In all experiments, it compared favorably against popular SDC approaches in terms of the confidentiality/utility tradeoff.

Due to its non-parametric and model free nature, the proposed approach might prove to be particularly effective in noisy and small datasets, which tend to be more challenging for approaches based on data modeling (which usually require good quality data in order to perform well). Quite importantly, such small/noisy datasets are likely the norm, rather than the exception, in academic research settings in health and social sciences, where preserving the confidentiality of the research subjects is key. (But, of course, the approach might also be a good option for large/high quality datasets, as well.)

While in this paper we focused on numerical variables, the approach can be directly applied to ordinal variables with high numbers of levels (which can be essentially treated as numerical variables). The approach might also be potentially extended to categorical variables with high numbers of levels, where it is possible to categorize the variables into a smaller number of levels (similarly, to the approach referred to as ``generalization", ``coarsening", or ``global recoding" in SDC), and then perform the restricted shuffles of the original values within the levels of the coarser categorical versions of the variables.

Finally, an important limitation of the proposed approach is that it cannot be directly extended to categorical or ordinal variables with low numbers of levels, where it is not possible to further coarse the variables into fewer levels. Possible research directions to handle this situation include hybrid approaches where SJPPDS is used to mask the continuous variables and an alternative approach is used to mask the categorical ones.

\beginappendix

\section*{Appendix}

\section{Experimental evaluation details}

\subsection{Disclosure risk metrics}

In our evaluations we adopt the following DR metrics:
\begin{enumerate}
\item The distance based record linkage (DBRL) metric~\cite{PagliucaSeri1999,DomingoFerrerTorra2001} is one of the most widely used methods for quantifying re-identification disclosure risk in SDC. It is implemented by first standardizing the variables in the data (to avoid scaling issues)~\cite{DomingoFerrerTorra2001}, and then computing the euclidean distances between each record in the masked dataset against all records in the original dataset. A masked record is then classified as ``linked" when the nearest record in the original dataset turns out to be the corresponding original record. The metric is then computed as the proportion of masked records that turn out to be linked to original records.
\item The rank interval distance (RID) metric~\cite{DomingoFerrerTorra2001} is a popular metric for measuring attribute disclosure risk. It corresponds to the proportion of original records inside a rank interval whose center is the corresponding masked record. The rank interval is computed as follows. Each variable in the masked data is independently ranked and a rank interval is defined around the value that the variable takes on each record, $r_{ij}$ (where $r_{ij}$ represent the rank of the $i$th record for $j$th variable). The rank interval is defined as $[r_{ij} - n \, p, r_{ij} + n \, p]$, so that the ranks of values within the interval differ by less than $p$ percent of the total number of records, $n$. A record in the original dataset is considered to be inside the rank interval of masked record $i$ if, for all variables $j$, it is inside the respective rank interval. The interval distance is then computed as the proportion of original records inside a rank interval. Following the recommendations in~\cite{DomingoFerrerTorra2001}, we report average RID values, obtained by averaging the RID for $p$ varying from 1\% to 10\% (in 1\% increments).
\item The standard deviation interval distance (SDID) metric~\cite{MateoSanzDomingoSebeFerrer2004} is another popular metric for measuring attribute disclosure risk. It is computed exactly as the RID metric, but with the exception that intervals are built around the raw values of the masked variables (rather than around their ranks), and the interval width is computed in terms of a percentage $p$ of the standard deviation of the variable (rather than in terms of a rank percentage).
\end{enumerate}

\subsection{Information loss metrics}

In our evaluations we adopt the following IL metrics:
\begin{enumerate}
\item Propensity score (PS) metric~\cite{Woo2009}. In the causal inference literature, the propensity score is defined as the probability of being assigned to treatment group, given the values of covariates~\cite{RosenbaumRubin1983}. When two large groups have similar distributions of propensity scores, the groups should have similar covariate distributions. In the context of SDC, one can stack the original and masked datasets, adding a variable ``treatment", set to 1 for masked records and to 0 for the original records, and then compute propensity scores (i.e., the probability of being a masked record) for all records in the stacked dataset. If the distribution of the propensity scores for the original records is similar to the distribution of propensity scores of the masked records, this means that the original data is similar to the masked data and the masking method has incurred a small amount of information loss. Following~\cite{Woo2009} we estimate the propensity scores using logistic regression using a second order polynomial on all variables and including all two-by-two interactions. (The propensity scores are simply the predicted probabilities of the logistic regression model.) The similarity of the propensity score distributions is computed as,
\begin{equation}
\mbox{PS} = \frac{1}{2 \, n} \, \sum_{i = 1}^{2 \, n} (p_i - 1/2)^2~,
\label{eq:propensity.score}
\end{equation}
where $2 n$ is the total number of records in the stacked dataset and $p_i$ is the propensity score for the $i$th record. Note that equation (\ref{eq:propensity.score}) will be maximal (assuming value 0.25) when $p_i$ is either 1 or 0 for all $i$ (in which case the original and masked datasets are completely distinguishable), and will be minimum (assuming value 0) when $p_i = 0.5$ for all $i$ (in which case the datasets are completely undistinguishable). Because PS is bounded in the interval [0, 0.25] we multiply its value by 4 in order to obtain an information loss metric with range in the [0, 1] interval.

\item Probabilistic information loss (PIL) metric~\cite{PilPaper2005} quantifies information loss from a probabilistically perspective based on the differences between statistics from the original and perturbed datasets. For any population parameter $\theta$ and corresponding sample statistic $\hat{\theta}$ in the original data, $\bfX$, the PIL approach computes the corresponding sample statistic $\hat{\theta}^\star$ on the masked dataset $\bfX^\star$, and measures information loss using the standardized discrepancy statistic, $Z = (\hat{\theta}^\star - \hat{\theta})/Var[\hat{\theta}^\star]^{1/2}$. Under the assumption that $Z$ converges to a $N(0, 1)$ distribution, the PIL approach quantifies the information loss w.r.t. $\theta$ through the probability,
\begin{equation}
\mbox{PIL}(\theta) = 2 \, P\left(0 \leq Z \leq |\hat{\theta}^\star - \hat{\theta}|/Var[\hat{\theta}^\star]^{\frac{1}{2}} \right)~,
\end{equation}
Note that $\mbox{PIL}(\theta)$ is 0 when there is no loss of information (i.e., $\hat{\theta}^\star = \hat{\theta}$), and increases towards 1 as the absolute discrepancy $|\hat{\theta}^\star - \hat{\theta}|$ increases. We are often interested in evaluating information loss across many distinct population parameters and following~\cite{PilPaper2005}, we compute the average PIL across means, variances, covariances, correlations, and sample quantiles.

\item Distance-based information loss metrics, including the mean absolute and mean squared error distances (MAE and MSE)~\cite{DomingoFerrerMateoSanzTorra2001,DomingoFerrerTorra2001b}, the bounded rank-based mean absolute and mean squared error distances (brMAE and brMSE)~\cite{ChaibubNeto2023}. For a dataset $\bfmX$ of dimension $n$ by $p$ the MAE and MSE metrics for measuring information loss between the original, $\bfmX$, and masked, $\bfmX^\star$, are given by~\cite{DomingoFerrerMateoSanzTorra2001,DomingoFerrerTorra2001b},
\begin{equation}
\mbox{MAE} = \frac{1}{n \, p} \, \sum_{i = 1}^{n} \sum_{j = 1}^{p} |x_{ij} - x_{ij}^\star|~,
\hspace{0.5cm}
\mbox{MSE} = \frac{1}{n \, p} \, \sum_{i = 1}^{n} \sum_{j = 1}^{p} (x_{ij} - x_{ij}^\star)^2~.
\end{equation}
One important drawback of these measures is that they are unbounded. This makes their comparison to the disclosure risk metrics (which are bounded in the [0, 1] interval) difficult. To circumvent the unbounded issue of the MAE and MSE metrics, reference~\cite{ChaibubNeto2023} has proposed bounded rank-based versions of these metrics (and showed that the bounded metrics tended to correlate well with their unbounded counterparts in practice). By working with the ranks, rather than with the raw numeric values of the data, it is possible to compute the maximum distance between a variable in the original and masked datasets (since the ranks of a variable are necessarily bounded between 1 and $n$). As described in~\cite{ChaibubNeto2023}, the bounded rank-based versions of the MAE and MSE metrics, denoted as brMAE and brMSE, are given by,
\begin{equation}
\mbox{brMAE} = \frac{1}{2 \, p \, \sum_{k = 1}^{K} (n - 2 \, k + 1)} \sum_{j = 1}^p \sum_{i = 1}^n |r_{ij} - r^{\star}_{ij}|~,
\end{equation}
\begin{equation}
\mbox{brMSE} = \frac{1}{2 \, p \, \sum_{k = 1}^{K} (n - 2 \, k + 1)^2} \sum_{j = 1}^p \sum_{i = 1}^n (r_{ij} - r^{\star}_{ij})^2~.
\end{equation}
where $K$ equals $n/2$ if $n$ is even and $(n-1)/2$ if $n$ is odd, and where $r_{ij}$ and $r^{\star}_{ij}$ represent, respectively, the ranks of the original, $x_{ij}$, and masked, $x^{\star}_{ij}$, raw data values. Similarly to the DR metrics, these metrics are also bounded in the [0, 1] interval. These metrics assume that the data have no ties (or that the ties have been broken by an approach, such as as random assignment of ranks to the tied values, that does not generate repeated values).
\end{enumerate}

\subsection{Simulated datasets details}

We performed 2 simulated data experiments. For the first, we generated data from a multivariate normal distribution, $\bfmX \sim N_p(\bfmu, \bfSigma)$, with distinct mean vectors and structured covariances matrices (with off-diagonal entries $\sigma_{ij} = \rho^{|i - j|}$, and diagonal entries $\sigma_{jj} = 1$). For each dataset, the mean values $\mu_j$, $j = 1, \ldots, p$, and correlation parameter $\rho$ were randomly sampled from $U(-3, 3)$ and $U(-0.8, 0.8)$ distributions, respectively. For the second simulated data experiment, we simulated correlated exponential random variables as follows. First, we simulate data from a multivariate normal random variable $\bfmZ \sim N_p(\bfzero, \bfSigma)$, then, for $j = 1, \ldots, p$, we compute the correlated uniform variables $U_j = \Phi(Z_j)$, and the correlated exponential random variables $X_j = G_{\lambda}^{-1}(U_j)$, where $\Phi$ and $G_{\lambda}$ represent, respectively, the cumulative distribution functions of standard normal and exponential (with rate $\lambda$) random variables. For each dataset, we randomly sample $\lambda$ and $\rho$ from $U(0.1, 10)$ and $U(-0.8, 0.8)$ distributions, respectively (and compute $\bfSigma$ as before). In each of the 2 experiments we generated 30 datasets, $\bfmX$, of dimension $n = 500$ by $p = 10$.

\subsection{Computation time benchmarking experiment details}

In all computation time benchmarking experiments, we simulated data from a $p$ dimensional multivariate normal distribution with mean vector $\bfmu = \bfzero$ and covariance matrix, $\bfSigma$, with off-diagonal entries $\sigma_{ij} = (-0.75)^{|i - j|}$ and diagonal entries $\sigma_{jj} = 1$. The tuning parameters were set to the following values: 100 for the SJPPDS methods; 7 for the microaggregation methods; 100\% for the noise addition approaches; 15\% for the rank-swapping method; and the original variable order for the cart approach.

\section{Time complexity of the SJPPDS algorithm}

Here we describe the computation complexity of the SJPPDS algorithm (Algorithm 3) based on the simplified version of the joint probability preserving data shuffling algorithm (Algorithm 2).

Starting with Algorithm 2, note that steps 1, 2, 3, 4, 6, 10 and 11 have complexity $O(1)$. Step 5 has complexity $O(n)$ since the Unique operation has linear complexity (and is applied to a vector of length $n$). The number of iterations of the for-loop in Step 7 is $n_c$ (since the number of unique values in the last column of $C$ will be equal to the number of classes/levels used in the discretization). Hence, the complexity of all steps involved in the for-loop is $O(n n_c)$. (Since: step 8 has complexity $O(n)$; step 9 has complexity lower than $O(n)$, as $idx$ has less than $n$ elements and the Shuffle operation has linear complexity; and step 10 has complexity $O(1)$.)  Hence, the total complexity of Algorithm 2 is $O(n n_c )$.

Now, moving to Algorithm 3, note that steps 1, 5, and 8 have complexity $O(1)$. Steps 2 and 6 have complexity $O(n p)$ (since the discretization operation is linear on the number of samples $n$ and we discretize the $p$ columns of the data). Steps 3 and 7 correspond to applications of Algorithm 2 and have complexity $O(n n_c)$. Since the discretization and joint probability preserving operations are repeated $p$ times, we have that the total complexity of Algorithm 3 is $O(p (n p + n n_c)) = O(n (p^2 + n_c p))$.

\section{Additional figures}

\begin{figure}[!h]
\centerline{\includegraphics[width=\linewidth]{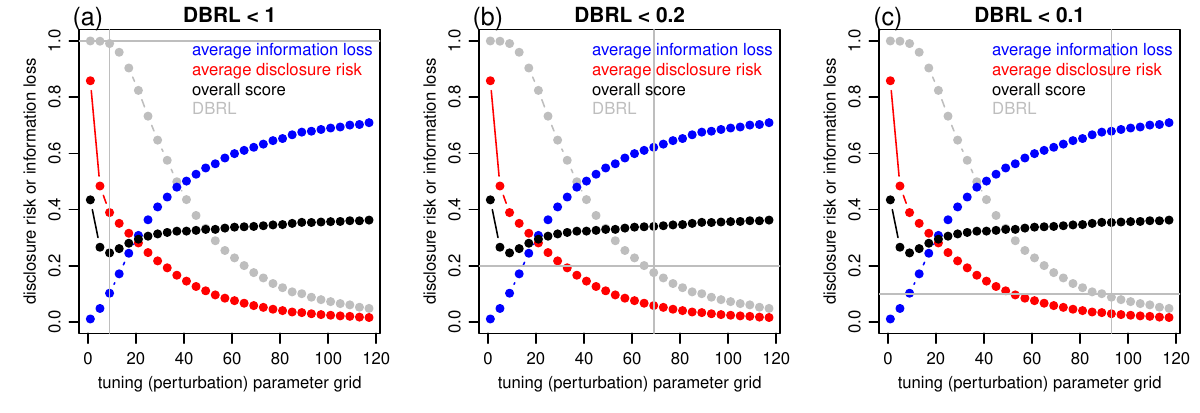}}
%\vskip -0.15in
\caption{Selection of best tuning/perturbation parameter values conditional on different DBRL thresholds. Panels a, b, and c report the tradeoff between information loss and disclosure risk across a grid of tuning/perturbation parameter values (which in this example corresponds to the percentage of additive noise). In all panels, the blue curve represents the average information loss score (equation \ref{eq:average.info.loss}), the red curve represents the average disclosure risk score (equation \ref{eq:average.disclosure.risk}), the black curve represents the overall score (equation \ref{eq:overall.score}), and the grey curve represents the DBRL metric. The grey horizontal line shows the selected DBRL threshold, while the grey vertical line shows the selected best parameter value. Panel a shows the results for DBRL threshold of 1. In this case the minimal overall score (intersection of the black curve and vertical grey line) is obtained at a perturbation parameter equal to 9\% of noise. Note, however, that the corresponding DBRL value (intersection of the grey curve and grey vertical line) is close to 1 (what is unacceptably high in practice). Panel b shows the results for DBRL threshold of 0.2. In this case, we only allow tuning/perturabtion parameter values for which the DBRL metric is lower than 0.2 (which is achieved at a tuning parameter value of 69\% of noise). Under this restriction we have that the lowest overall score (black curve) is also achieved at 69\% of noise. Panel c shows analogous results for DBRL threshold of 0.1, in which case the best tradeoff is achieved at 93\% of noise.}
\label{fig:dbrl.thresholds}
%\vskip -0.1in
\end{figure}

\begin{figure}[!h]
\centerline{\includegraphics[width=\linewidth]{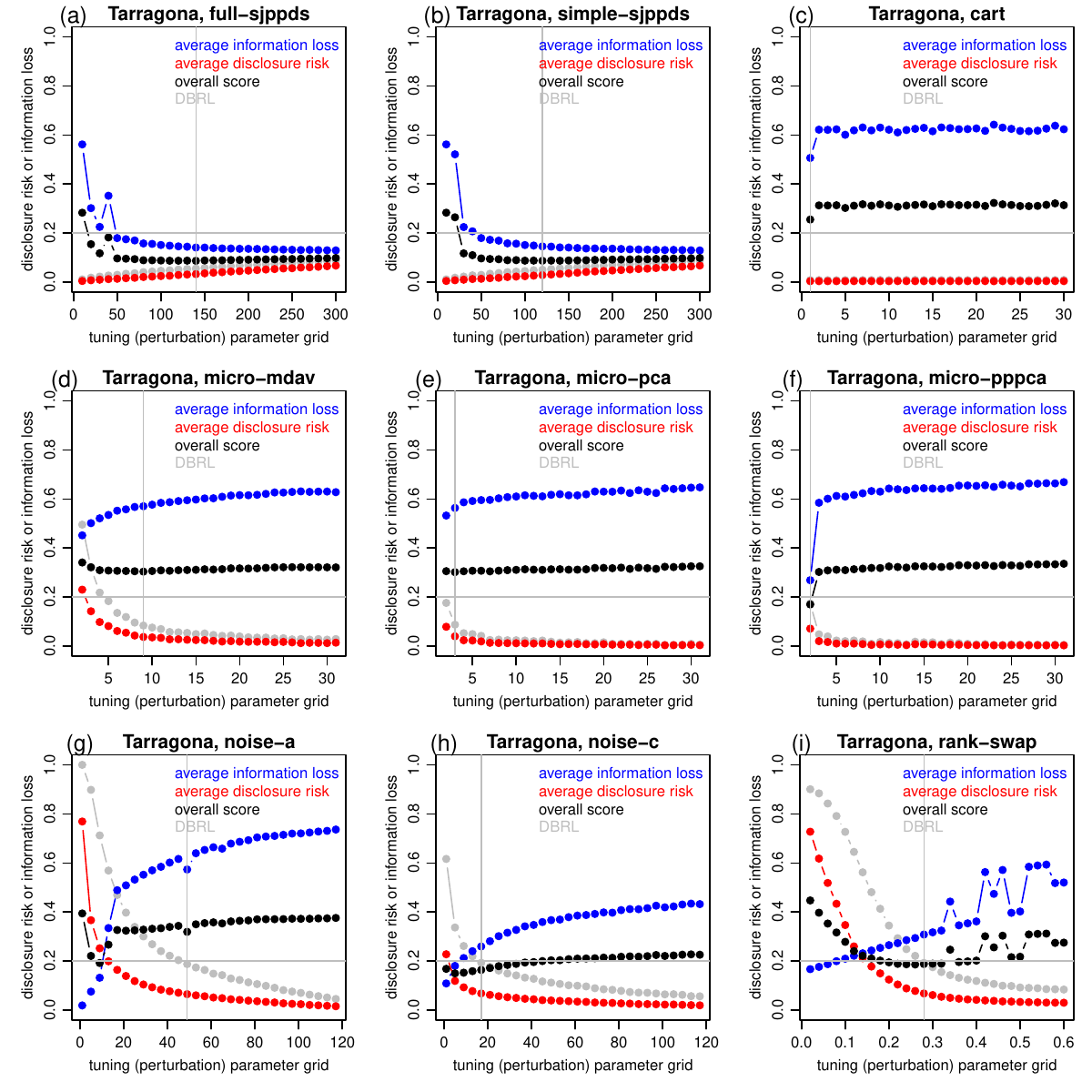}}
%\vskip -0.15in
\caption{Tradeoff between information loss and disclosure risk in the Tarragona dataset.}
\label{fig:tradeoff.tarragona}
%\vskip -0.1in
\end{figure}

\end{document}